\documentclass[aps,pre,twocolumn,nofootinbib]{revtex4-1}

\usepackage{graphicx}
\usepackage{amsmath}
\usepackage{color}
\usepackage{epstopdf}

	\usepackage[colorlinks,hyperindex]{hyperref}
 \usepackage{hyperref}
	\hypersetup
	{
		colorlinks,%
		citecolor=blue,%
		linkcolor=blue,%
		urlcolor=black,%
	}

\begin{document}

\title{Properties of compressible elastica from
  relativistic analogy}

\author{Oz Oshri} 
\email{ozzoshri@tau.ac.il} 
\affiliation{Raymond \& Beverly Sackler School of Physics \& Astronomy, Tel Aviv University, Tel Aviv 6997801, Israel}

\author{Haim Diamant} 
\email{hdiamant@tau.ac.il} 
\affiliation{Raymond \& Beverly Sackler School of Chemistry, Tel Aviv University, Tel Aviv 6997801, Israel}

\date{\today}

\begin{abstract} 
Kirchhoff's kinetic analogy relates the deformation of an
incompressible elastic rod to the classical dynamics of rigid body
rotation. We extend the analogy to compressible filaments and find
that the extension is similar to the introduction of relativistic
effects into the dynamical system. The extended analogy reveals a
surprising symmetry in the deformations of compressible elastica. In addition, we use known results for the buckling
of compressible elastica to derive the explicit solution for the
motion of a relativistic nonlinear pendulum. We discuss cases where the extended Kirchhoff analogy may be useful for the study of other soft matter systems.
\end{abstract}
\maketitle

Analogies to dynamical problems have been used to simplify the physics
of various condensed-matter systems, ranging from the deformation of
elastic bodies to the order-parameter profile across an interface
between coexisting phases. A particularly well known example is
Kirchhoff's kinetic analogy \cite{love}. In this theory the
three-dimensional (3D) deformation of a slender elastic rod is reduced
to the bending deformation of an incompressible curve, representing
the mid-axis of the rod. This problem, in turn, is analogous to the
dynamics of a rigid body rotating about a fixed point, where the
distance along the curve and its local curvature are analogous,
respectively, to time and angular velocity. When the filament is
confined to a two-dimensional (2D) plane (the celebrated Euler
elastica \cite{euler}), the equation of equilibrium coincides with the
equation of motion of a physical pendulum \cite{love,nizette}.


In the examples above the elastic system was reduced to an
indefinitely thin, incompressible body, whose equilibrium shape
follows the trajectory of a classical dynamical system. In the present
work we show that relaxing the incompressibility constraint introduces terms akin
to relativistic corrections to classical dynamics. Within this
analogy, the compression modulus, $Y$, plays the role of the
relativistic particle's rest mass, and the bendability parameter
$(Y/B)^{1/2}\equiv h^{-1}$, where $B$ is the bending modulus, is
analogous to the speed of light. The limit of an incompressible rod
($h\rightarrow 0$) corresponds to the nonrelativistic limit.

\begin{figure}[!tbh]
\vspace{0.7cm}
\includegraphics[width=\columnwidth]{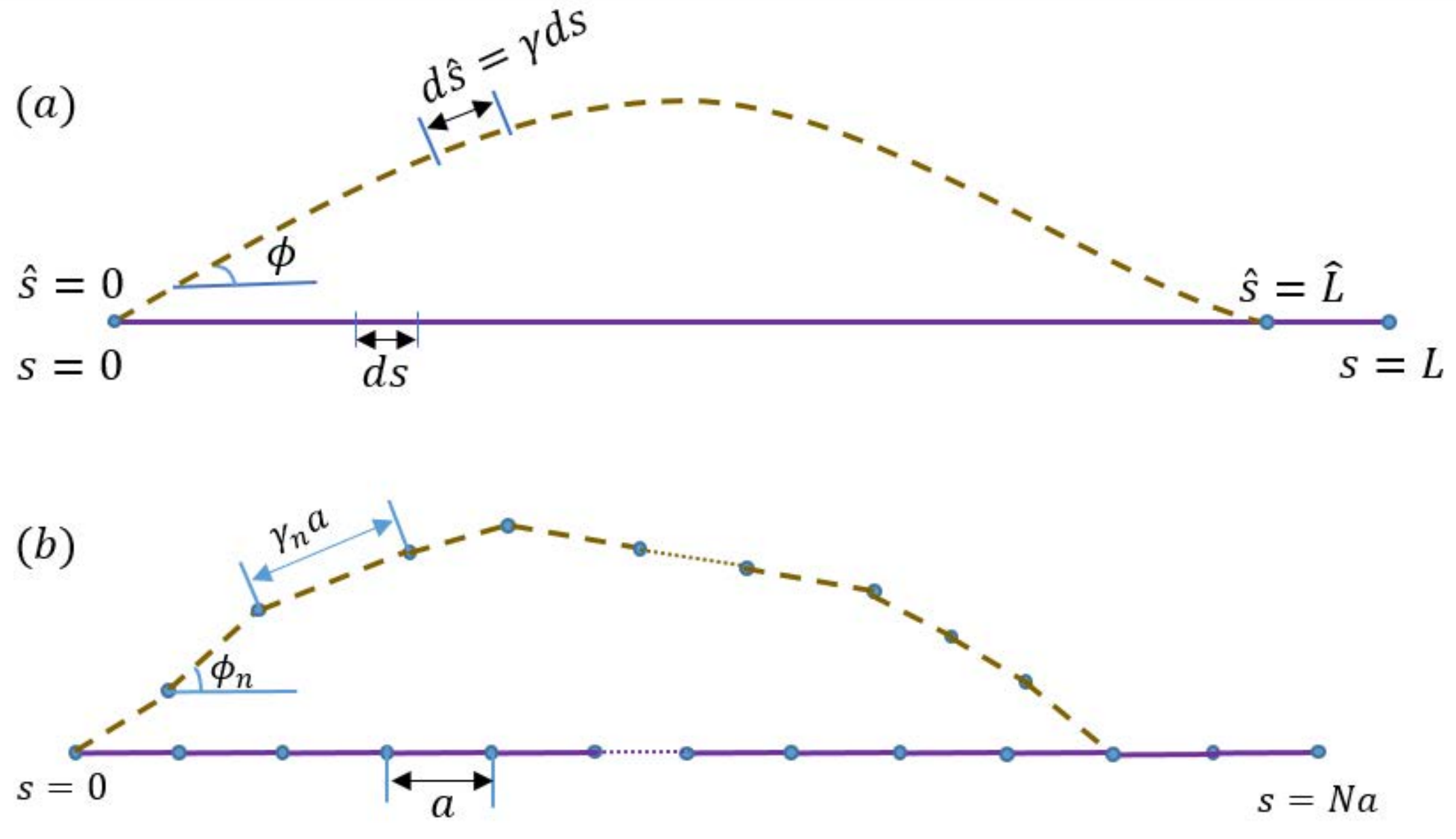}
\caption{(a) Deformed and undeformed configurations. The straight line
  (solid, purple) represents a relaxed rod of initial length $L$, and
  the curved line (dashed, gold) represents a deformed rod of total
  length $\hat{L}$. Their arclength parameters are $s$ and $\hat{s}$
  respectively. The local in-plane and out-of-plane deformations are
  respectively accounted for by the compression field
  $\gamma=d\hat{s}/ds$ and the angle $\phi(s)$. (b) An illustration of
  the discrete model. The initial, zero-energy configuration (solid,
  purple) consists of $N$ rigid bonds of rest length $a$ and zero
  joint angles. A higher-energy state (dashed, gold) is obtained by
  compression of each bond, $a\rightarrow\gamma_n a$, and/or a change
  of each joint angle, $\phi_n$. The continuous limit of this model
  yields the picture presented in panel (a).}
\label{fig01}
\end{figure}

Despite the relevance to real systems, including compressible fluid
membranes \cite{Markus}, there have not been many studies of
compressible elastica (see \cite{magnusson} and references therein). Following these works,
we consider the 2D deformation of a compressible filament, represented
by a planar curve of relaxed length $L$. The same model applies to
thin elastic sheets, as well as fluid membranes
\cite{vassilev}, provided that they are deformed along
a single direction. The deformation away from the flat, stress-free
state is parametrized by the angle $\phi(s)$ and compressive strain
$\gamma(s)$, as functions of the relaxed arclength $s$ along the
curve, $s\in[0,L]$; see Fig.~\ref{fig01}(a).  We denote the compressed
arclength by $\hat{s}$, such that $\gamma=d\hat{s}/ds$ and the total
deformed length is
$\hat{L}=\int_{0}^{\hat{L}}d\hat{s}=\int_{0}^{L}\gamma(s)ds$.

To obtain the energy functional which keeps the bending and
compression contributions independent, it is instructive to start from
a discrete model (see Fig.~\ref{fig01}(b)). Consider a chain of $N$
jointly connected compressible rigid bonds of rest length $a$. The
chain's configuration is parametrized by a set of bond strains
$\gamma_n$ ($1\leq n\leq N$) and a set of joint angles $\phi_n$
($1\leq n\leq N-1$).  The energy of a given configuration has a
compression contribution, $E_c=(Y/2a)\sum_{n=1}^{N}(\gamma_n a-a)^2$,
and a bending contribution,
$E_b=(B/2a)\sum_{n=1}^{N-1}(\phi_{n+1}-\phi_n)^2$. For the sake of the
analogy we add a potential energy of the form,
$E_p=a\sum_{n=1}^{N}\gamma_n V(\phi_n)$, where $V(\phi)$ is an
angle-dependent potential. This choice of
potential energy is not artificial. For example, if we include an
external work on the chain, coupling the force $P$ exerted on the
boundaries with the chain's projected length, we have
$E_p=Pa\sum_{n=1}^{N}\gamma_n\cos\phi_n$, i.e. $V(\phi)=P\cos\phi$.
We now take the continuum limit, $N\rightarrow \infty$ and
$a\rightarrow 0$ such that $Na\rightarrow L$ and $na \rightarrow s$,
getting,
\begin{eqnarray}
&E&[\phi(s),\gamma(s)]=E_b+E_c+E_p \nonumber  \\  
&=&\int_{0}^{L}ds\left[\frac{B}{2}\left(\frac{d\phi}{ds}\right)^2+\frac{Y}{2}(\gamma-1)^2+\gamma V(\phi)\right].
\label{energyE}
\end{eqnarray}       

To obtain the equations of equilibrium for the filament one should
minimize $E$ with respect to $\gamma(s)$ and $\phi(s)$.
This gives the two equations,
\begin{eqnarray}
&&Y(\gamma-1)+V=0 \label{eqn_gamma},\\
&&B \frac{d^2\phi}{ds^2}-\gamma \frac{dV}{d\phi}=0 \label{eqn_phi}.
\end{eqnarray}
First integration yields,
\begin{equation}
\mathcal{H}=\frac{B}{2}\left(\frac{d\phi}{ds}\right)^2-\frac{Y}{2}(\gamma-1)^2-\gamma V=\text{const},
\label{constantH}
\end{equation} 
which depends on the boundary conditions at $s=0,L$. 

The mathematical analogy to relativistic dynamics is revealed once we transform from the relaxed arclength, $s$, to the compressed one, $\hat{s}$, and redefine the potential as $\bar{V}=-V$. Equation~(\ref{eqn_phi}) then turns into
\begin{equation}
B\frac{d}{d\hat{s}}\left(\gamma \frac{d\phi}{d\hat{s}}\right)+\frac{d\bar{V}}{d\phi}=0.
\label{relativisticEq}
\end{equation}
This equation is analogous to the Euler-Lagrange equation for the one-dimensional
motion of a relativistic particle \cite{landau}, provided that $\gamma(\hat{s})$ coincides with the Lorentz factor. To obtain $\gamma$ we use
eqn~(\ref{constantH}) to eliminate $V$ in eqn~(\ref{eqn_gamma}), and
transform to $\hat{s}$, which gives,
\begin{equation}
\gamma=\frac{\sqrt{1+2\mathcal{H}/Y}}{\sqrt{1+h^2(d\phi/d\hat{s})^2}}.
\label{gamma_el}
\end{equation}
This expression indeed resembles the Lorentz factor up to the constant prefactor, $\gamma_0\equiv \sqrt{1+2\mathcal{H}/Y}$. Absorbing this prefactor in the potential $\bar{V}$ makes eqn~(\ref{relativisticEq}) identical to the relativistic one, with the correct Lorentz factor. This completes the analogy. The mapping between the variables
and parameters of these two systems is summarized in Table
\ref{comperison-table}.

\begin{table}[ht]
\caption{Mapping between the parameters of compressible elastica and relativistic dynamics}
\begin{center}
\begin{tabular}{| l | l |}
\hline
{\bf Elasticity} & {\bf Relativity} \\
\hline\hline
relaxed arclength\footnote{\label{gamma-constant} More accurately, the analogue of $\tau$ is $\gamma_0 s$, where $\gamma_0$ is a constant prefactor (see text).  }, $s$ & proper time, $\tau$ \\
\hline
compressed arclength, $\hat{s}$ & laboratory time, $t$ \\
\hline
tangent angle, $\phi$ & angle coordinate, $\phi$ \\
\hline
curvature, $\kappa=d\phi/d\hat{s}$ & angular velocity, $\omega=d\phi/dt$ \\
\hline
compression field$^{\rm\ref{gamma-constant}}$, $\gamma=d\hat{s}/ds$ & Lorentz factor, $\gamma=dt/d\tau$ \\
\hline
compression modulus, $Y$ & minus the rest energy, $-mc^2$ \\
\hline
bending modulus, $B$ & moment of inertia, $m\ell^2$ \\
\hline
bendability, $h^{-1}=\sqrt{Y/B}$ & speed of light, $c$\\
\hline
\end{tabular}
\end{center}
\label{comperison-table}
\end{table}

\begin{figure}[!tbh]
\vspace{0.7cm}
\includegraphics[width=70mm,scale=0.4]{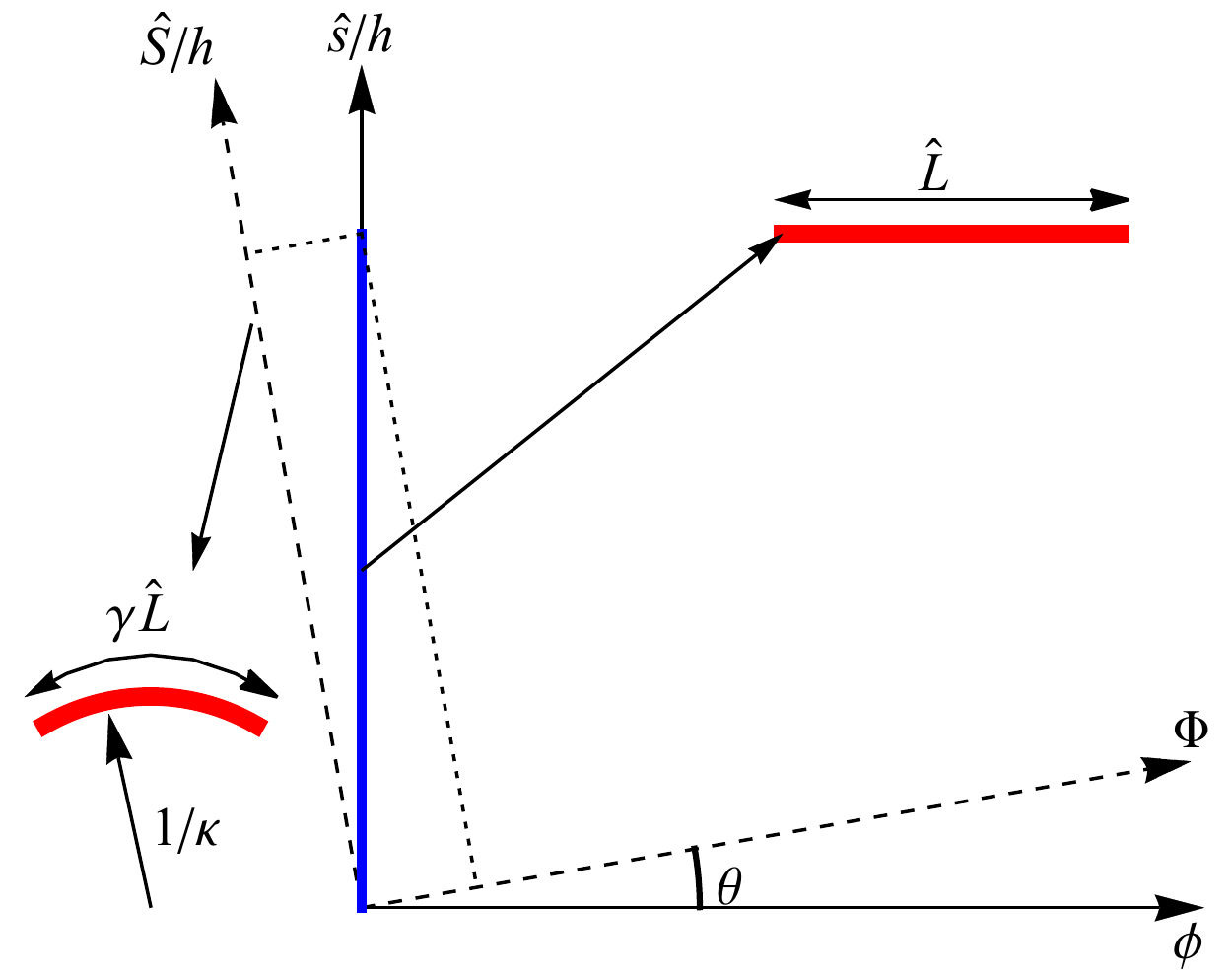}
\caption{Rigid rotation of the coordinate system
  $(\hat{s}/h,\phi)\rightarrow (\hat{S}/h,\Phi)$ by an angle
  $\theta$. A flat elastic filament of length $\hat{L}$ is described
  by the blue line. As viewed from the new frame, the total length is
  compressed by a factor $\gamma$, and the angle coordinate, $\Phi$,
  is linearly changing, resulting in a cylindrical compressed shape of
  radius $1/\kappa$ and length $\gamma\hat{L}$.}
\label{LorentzRotation}
\end{figure}

We now use the mapping to identify a new symmetry of
compressible elastica. First, recall that the equation of equilibrium
of {\it incompressible} elastica is invariant to the addition of a
constant curvature $\kappa$, making a flat configuration
cylindrical. The kinetic analogue of this symmetry is the Galilean
invariance of the equation of motion of a classical free particle to a
boost by a constant velocity. We are after the corresponding symmetry
for a {\it compressible} elastic filament. It is natural to try the
analogue of a Lorentz boost. To make the connection to a free particle
we first remove the potential energy from eqn~(\ref{relativisticEq}),
$\bar{V}(\phi)=0$, and then divide by $B\gamma_0$. The resulting equation of elastic equilibrium,
\begin{equation}
\frac{d}{d\hat{s}}\left(\frac{1}{\sqrt{1+h^2(d\phi/d\hat{s})^2}}\frac{d\phi}{d\hat{s}}\right)=0,
\label{free-rel}
\end{equation}
 is identical to the equation of motion of a free relativistic particle.   
With the Lorentz boost in mind, we immediately identify the
transformation of coordinates $(\hat{s}/h,\phi)\rightarrow
(\hat{S}/h,\Phi)$, which leaves eqn (\ref{free-rel})
unchanged:
\begin{eqnarray}
\begin{pmatrix}
\hat{s}/h \\ \phi
\end{pmatrix}=
\begin{pmatrix}
\gamma & -\gamma h \kappa \\ \gamma h \kappa & \gamma
\end{pmatrix} 
\begin{pmatrix}
\hat{S}/h \\ \Phi
\end{pmatrix},
\label{LorentzBoost}
\end{eqnarray}
where $\kappa=\text{const}$, and
$\gamma=1/\sqrt{1+h^2\kappa^2}$. The transformation is a rotation of
the material coordinates, $(\hat{s}/h,\phi)$, by an angle
$\theta=\sin^{-1}(\gamma h \kappa)$, turning a flat, relaxed
configuration into a cylindrical, compressed one; see
Fig.~\ref{LorentzRotation}. The rotation leaves the line element on the material-coordinates plane unchanged, $d\hat{s}^2/h^2+d\phi^2=d\hat{S}^2/h^2+d\Phi^2$.
This reparametrization invariance defines a continuous family of solutions to the equation of equilibrium, containing all the cylindrical, compressed configurations of the filament. These configurations satisfy local mechanical equilibrium but differ in their total energy. (Obviously, the flat relaxed configuration always has the lowest energy.) The actual equilibrium configuration for a given problem is selected by boundary conditions.
To our knowledge, this symmetry of
compressible filaments has not been recognized before.

An important physical distinction should be made, however, between the two
sides of the analogy. In the elastic problem there is a unique zero-energy state, and one naturally specifies the length
$L$ and the boundary conditions with respect to that relaxed state; the compressed
length $\hat{L}$ is determined by minimization. By contrast, in relativity there is no preferred reference frame, and the duration of the experiment and the boundary
conditions are given in the laboratory frame, not the proper one. This difference leads to the extra prefactor $\gamma_0$ above. Thanks to this prefactor
the elastic strain can be
either compressive or dilative, whereas the relativistic Lorentz factor must be
dilative.\footnote{If we formulated the elastic problem and its boundary conditions, unphysically, based on the ultimate deformed length $\hat{L}$, the information on the unique reference state would be lost, and the two problems, the elastic and relativistic, would become identical. This can be readily shown by transforming eqn (\ref{energyE}) to $\hat{s}$ and minimizing with respect to $\gamma$.} Note that, because of the minus sign in the mapping $Y\rightarrow -mc^2$, time dilation corresponds to elastic compression. 


Despite the physical difference, by absorbing the factor $\gamma_0$ in the
potential, the two problems become mathematically equivalent.
Thus, a known solution in one system can be readily
transferred to the other. We
shall now use known results for the buckling of compressible elastica to
write down an explicit solution to the equation of motion of a relativistic nonlinear pendulum. (A solution to this dynamical system was
previously given only in implicit parametric form \cite{denis}.)

To do so, we should specialize to the potential $V(\phi)=P\cos\phi$.
Substitution of $V$ in eqn~(\ref{eqn_gamma}) gives
$\gamma=1-(P/Y)\cos\phi$. These expressions for $V$ and $\gamma$ turn
eqn~(\ref{eqn_phi}) into
\begin{equation}
B\frac{d^2\phi}{ds^2}+P\sin\phi-\frac{P^2}{Y}\sin\phi\cos\phi=0,
\label{compressible_elastica}
\end{equation} 
which is the equation of equilibrium for a compressible elastica under
a uniaxial force $P$ \cite{magnusson,humber}. The exact solution to
eqn~(\ref{compressible_elastica}) was derived in
ref.~\cite{magnusson}. For hinged boundary conditions, $d\phi/ds=0$ at
$s=0,L$, it is given by
\begin{equation}
\phi(s)=2\sin^{-1}\left[q\frac{(1-(q k h)^2/\gamma_0 )^{1/2}\,\text{cd}\left(\sqrt{\gamma_0 }ks,m\right)}{\left[1-(q k h)^2 /\gamma_0 \,\text{cd}^2\left(\sqrt{\gamma_0 }ks,m\right)\right]^{1/2}}\right].
\label{PP_solution}
\end{equation}
Hereafter we use the conventional symbols for the various elliptic functions, $\text{cd},\text{am},K$ and $\Pi$, as defined in ref.~\cite{Abramowitz}. In addition, $k\equiv\sqrt{P/B}$ is the wavenumber of the
buckled {\em linear} elastica. We have defined three additional
parameters which depend on the boundary angle,
$\phi_0\equiv\phi(s=0)$:
\begin{eqnarray}
q&\equiv&\sin(\phi_0/2) \nonumber \\
\gamma_0&=&1-(k h)^2\cos\phi_0 \nonumber\\ 
m&=&q^2\left[1+(q k h)^2\right]/\gamma_0.
\label{mANDxi}
\end{eqnarray} 
The parameter $\gamma_0=\gamma(s=0)$ gives the boundary strain, which coincides with $\gamma_0$ as defined above. The
actual wavelength (periodicity) $\lambda$ of the nonlinear buckled
elastica is
\begin{equation}
  \lambda = 4K(m)/(\sqrt{\gamma_0}k).
\label{lambda}
\end{equation}
The total rod length matches a half wavelength,
$L=\lambda/2$. This expression relates $\phi_0$ to the force $P$ and
the system parameters via eqn~(\ref{mANDxi}). The total deformed
length is given by $\hat{L}=\int_0^{L}[1-(k h)^2\cos\phi]ds$,
yielding,
\begin{equation}
\hat{L}/L=1-(k h)^2\left[2(1-q^2)\frac{\Pi(q^2,m)}{K(m)}-1\right].
\label{period-hatL}
\end{equation}
It is readily shown that the three known solutions
for incompressible nonlinear elastica ($h\rightarrow 0$) \cite{love},
compressible linear elastica ($\phi_0\ll 1$) \cite{magnusson}, and
incompressible linear elastica ($h\rightarrow 0, \phi_0 \ll 1$)
\cite{love}, are obtained from eqn~(\ref{PP_solution}) in the
respective limits.

To apply the analogy, we first transform
eqn~(\ref{compressible_elastica}) to the deformed arclength $\hat{s}$
using equations~(\ref{eqn_gamma}),~(\ref{constantH}) and (\ref{gamma_el}).
This leads to
\begin{equation}
\frac{d}{d\hat{s}}\left(\frac{1}{\sqrt{1+h^2(d\phi/d\hat{s})^2}}
\frac{d\phi}{d\hat{s}}\right)+\frac{k^2}{\gamma_0}\sin\phi=0.
\label{compressible-shat-equation}
\end{equation}
Next, substituting
\[
\{h^2,k^2/\gamma_0 \}\rightarrow \{-\ell^2/c^2,\omega^2\},
\]
where $\omega$ is the natural frequency of the corresponding linear
pendulum and $\ell$ the pendulum length, the equation of motion for a
relativistic pendulum \cite{moreau} readily follows:
\begin{equation}
\frac{d}{dt}\left(\frac{1}{\sqrt{1-(\ell^2/c^2)(d\phi/dt)^2}}
\frac{d\phi}{dt}\right)+\omega^2 \sin\phi=0.
\label{PP_dynamic}
\end{equation}
From eqn~\ (\ref{PP_solution}) we then immediately write down the
explicit solution for the pendulum motion in terms of its proper time,
\begin{equation}
\phi(\tau)=2\sin^{-1}\left[q\frac{[1+(q\omega\ell/c)^2]^{1/2}\,\text{cd}
\left(\omega\tau,m\right)}{[1+(q\omega\ell/c)^2\,\text{cd}^2
\left(\omega\tau,m\right)]^{1/2}}\right],
\label{PP_solution_proper}
\end{equation} 
where, from eqn~\ (\ref{mANDxi}), $q=\sin(\phi_0/2)$ is related to the
initial angle of the pendulum, and
$m=q^2[1-(\omega\ell/c)^2(1-q^2)]$. Similarly, the proper period of the
pendulum is obtained by analogy to $\gamma_0\lambda$ of
eqn~\ (\ref{lambda}),
\begin{equation}
  T_{\tau} = 4K(m)/\omega,
\end{equation}
and the laboratory period from $\hat{L}/(\gamma_0L)$ of
eqn~\ (\ref{period-hatL}),
\begin{equation}
  T_t/T_{\tau} = 1+2(\omega\ell/c)^2(1-q^2)
  \left[\frac{\Pi(q^2,m)}{K(m)}-1\right].
\end{equation}
Finally, we can transform the solution (\ref{PP_solution_proper}) to
laboratory time through
\begin{equation}
  t = \tau+\frac{2(1-q^2)}{\omega}(\omega\ell/c)^2\left[
 \Pi(q^2,\text{am}(\omega\tau,m),m)-\omega\tau\right],
\label{totau}
\end{equation}

\begin{figure}[!tbh]
\vspace{0.1cm}
\centerline{\resizebox{0.5\textwidth}{!}
{\includegraphics{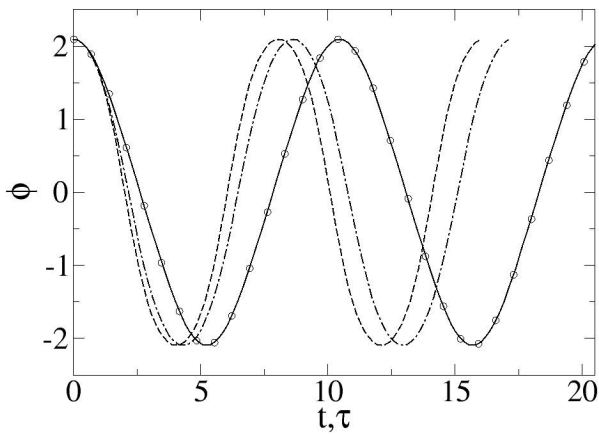}}}
\caption[]{Pendulum angle as a function of proper time
  (eqn~(\ref{PP_solution_proper}), dashed line) and laboratory time
  (equations~(\ref{PP_solution_proper}) and (\ref{totau}), solid line). The
  chosen parameters, $\phi_0=2\pi/3$, $\omega=1$, $(\omega\ell/c)^2=0.5$,
  correspond to a highly relativistic pendulum resulting in a
  significant time dilation. Numerical solution of the equation of
  motion (eqn~(\ref{PP_dynamic}), circles) and the case of a
  nonrelativistic pendulum ($c\rightarrow\infty$, dash-dotted) are
  plotted for comparison.}
\label{phiot}
\end{figure}

In Fig.~\ref{phiot} we show the
resulting motion of the pendulum in both proper and
laboratory times. The dilation of the period of the relativistic pendulum (solid curve), compared  to the nonrelativistic one (dash-dotted), illustrates the compression of the buckling wavelength in the compressible elastica compared to the incompressible case.  

We note that a similar factor to the elastic $\gamma_0$ is
encountered in certain relativistic scenarios \cite{denis,anderson}. We
note also that the usage of the original strain field 
to derive the elastic $\gamma$ 
looks similar in spirit to the ``einbein'' used
to simplify the analysis of actions in different physical contexts
\cite{GSW}. Unlike those analyses, however, $\gamma$ here is an actual
physical strain, the choice of which is not free but 
dictated by elasticity.

The extended Kirchhoff analogy presented in this Communication should be useful for studying various soft matter systems. In principle, compressibility corrections are negligible for sufficiently slender (and therefore bendable) objects. There are several indications, however, that the compressible-elastica model may be qualitatively predictive beyond its strict limits of validity. For example, it produces negative compressibility (decrease of the force $P$ with increasing confinement) for bendability $h^{-1}$ of order 1 \cite{magnusson}. Although the model is invalid for such values of $h^{-1}$, this result is in qualitative agreement with recent experimental and numerical results for buckled beams \cite{Coulais}. 
In ref.~\cite{Markus} Diggins {\it et al.} studied the buckling of fluid membranes under uniaxial confinement using molecular dynamics simulations. To improve the analysis of the simulation data beyond the incompressible Euler elastica, they used, on purely empirical grounds, a model which is very similar to the one presented here, obtaining good fits to the numerical results.  
The reason for the unexpected success of the compressible-elastica model in these two examples is unclear at present. Compressibility is essential in the case of two-dimensional deformations with non-zero Gaussian curvature. The present 1D model, as far as we know, has no applicability in such cases. It is relevant, however, to deformations with cylindrical \cite{Stoop,Lee} or conical \cite{Cerda} symmetries.

We have shown above how the buckling of a compressible filament can be studied as the oscillation of a relativistic nonlinear pendulum. 
Here are a few suggestions for future extensions. The instability of a flat compressible rod or sheet toward
curved configurations is connected to the emergence of oscillations
from rest in the parametric resonance of a relativistic oscillator. A
change in the compression modulus of an elastic system (e.g., as a
result of a temperature change) is likened to a change in a particle's
rest mass. Lastly, the analogy can be extended to three
dimensions, where the deformation of a compressible filament in 3D becomes analogous to a relativistic rotor, as outlined in Appendix~\ref{3Danalogy}.

\begin{acknowledgments}
We thank Markus Deserno for sharing unpublished simulation results
with us and for helpful discussions. This work has been supported in
part by the Israel Science Foundation (Grant No. 164/14).
\end{acknowledgments}

\appendix 

\section{Analogy between compressible 3D Kirchhoff's rod and relativistic rotor}

\label{3Danalogy}

In this appendix we derive the equations of equilibrium for a 3D compressible elastic rod under the assumptions of Kirchhoff's model \cite{dill,nizette}. It is shown that these equations are mathematically analogous to the 3D Euler's equations of rigid-body rotation, where the non-relativistic angular momentum, ${\bf L}$, is replaced by, $\gamma{\bf L}$, $\gamma$ being the Lorentz factor. This analogy is derived while keeping in mind the known difficulties in the latter, relativistic problem \cite{rizzi,Rynasiewicz}.

Following the formulation in ref.~\cite{lafortune}, we consider a space curve described by the position vector, ${\bf R}(s)$, where $s$ is the arclength parameter of the relaxed configuration. Defining $\hat{s}$ as the arclength parameter of the deformed configuration, the unit tangent vector to ${\bf R}(s)$ is given by, ${\bf d}_3=d{\bf R}/d\hat{s}$. In addition, let $\{{\bf d}_1,{\bf d}_2\}$ be a pair of unit vectors perpendicular to ${\bf d}_3$ and parallel to the principal axes of the filament's cross-section. The triad $\{{\bf d}_1,{\bf d}_2,{\bf d}_3\}$ form a co-moving coordinate frame attached to the rod's mid-axis. These vectors satisfy the relations,
\begin{equation}
\frac{d{\bf d}_i}{ds}=\gamma\pmb{\kappa}\times{\bf d}_i,
\label{FS3D}
\end{equation}
where $i=1,2,3$, $\gamma=d\hat{s}/ds$ is the strain field, and $\pmb{\kappa}=\kappa_1{\bf d}_1+\kappa_2{\bf d}_2+\kappa_3{\bf d}_3$ is the curvature vector. The local bending moment is given by,
\begin{equation}
{\bf M}(s)=B_1 (\gamma \kappa_1){\bf d}_1+B_2 (\gamma \kappa_2){\bf d}_2+B_3 (\gamma \kappa_3){\bf d}_3,
\label{M-compressible}
\end{equation}
where $B_i$ are bending rigidities. 

In correspondence with eqn~(\ref{energyE}), the energy functional of a 3D elastic filament, confined by a boundary constant force ${\bf P}$, is given by $E=\int_0^L e[\pmb{\kappa}(s),\gamma(s)] ds$, where 
\begin{equation}
e[\pmb{\kappa}(s),\gamma(s)]=\frac{1}{2}{\bf M}\cdot(\gamma \pmb{\kappa})+\frac{Y}{2}(\gamma-1)^2+\gamma {\bf d}_3\cdot {\bf P}.
\label{energyE3D}
\end{equation}
The appearance of the $\gamma \pmb{\kappa}$ term in the bending energy is a consequence of the requirement to keep bending and compression contributions independent. Minimizing eqn~(\ref{energyE3D}) with respect to $\gamma$ and $\kappa_i$ (keeping in mind that $\kappa_i$ are not independent) gives \cite{lafortune}, 
\begin{eqnarray}
\frac{d{\bf M}}{d\hat{s}}-{\bf d}_3\times{\bf P}&=&0, \label{eqn_phi3D} \\
Y(\gamma-1)+{\bf d}_3\cdot {\bf P}&=&0 \label{eqn_gamma3D}.
\end{eqnarray}

In the incompressible limit eqn~(\ref{eqn_gamma3D}) is redundant and equations~(\ref{eqn_phi3D}) become analogous to the non-relativistic Euler equations of a 3D rigid body, fixed at one point and rotating under the influence of an external force (such as gravity) \cite[p.~200]{goldstein}. In this analogy, the bending moment takes the role of angular momentum, ${\bf M}\leftrightarrow {\bf L}$, and the boundary force is analogous to an external torque, ${\bf d}_3\times{\bf P}\leftrightarrow {\bf N}$. 
Turning on compressibility effects, we have by eqn~(\ref{M-compressible}) that ${\bf M}\rightarrow \gamma {\bf M}$. Thus, it is left to show that $\gamma$ coincides with the Lorentz factor.
First integration of equations~(\ref{eqn_phi3D}) and (\ref{eqn_gamma3D}) gives,
\begin{equation}
\mathcal{H}=\frac{1}{2}{\bf M}\cdot(\gamma \pmb{\kappa})-\frac{Y}{2}(\gamma-1)^2-\gamma {\bf d}_3\cdot {\bf P}=\text{const}.
\label{constantH3D}
\end{equation}
Eliminating ${\bf d}_3\cdot {\bf P}$ from eqn~(\ref{constantH3D}) and substituting in eqn~(\ref{eqn_gamma3D}) gives,
\begin{equation}
\gamma=\frac{\sqrt{1+2\mathcal{H}/Y}}{\sqrt{1+\frac{B_1}{Y}\kappa_1^2+\frac{B_2}{Y}\kappa_2^2+\frac{B_3}{Y}\kappa_3^2}}.
\label{gamma_el3D}
\end{equation}
This expression indeed resembles the Lorentz factor
up to the constant prefactor, $\gamma_0\equiv \sqrt{1+2\mathcal{H}/Y}$, which appears also in the 2D problem, and which can be absorbed in the force ${\bf P}$, (see discussion in the main text).


\end{document}